\documentstyle[prd,aps,epsf,floats,amsfonts,amssymb,amsmath]{revtex}
\newcommand{\be}{\begin{equation}}\newcommand{\ee}{\end{equation}}
\newcommand{\bea}{\begin{eqnarray}}\newcommand{\eea}{\end{eqnarray}}
\newcommand{\beaa}{\begin{eqnarray}}\newcommand{\eeaa}{\end{eqnarray}}
\newcommand{\ba}{\begin{array}}\newcommand{\ea}{\end{array}}
\newcommand{\bit}{\begin{itemize}}\newcommand{\eit}{\end{itemize}}
\newcommand{\ben}{\begin{enumerate}}\newcommand{\een}{\end{enumerate}}
\def\lab{\label}\def\lan{\langle}\def\lar{\leftarrow}
\def\lf{\left}\def\lrar{\leftrightarrow}
\def\non{\nonumber}\def\pa{\partial}
\def\ran{\rangle}\def\rar{\rightarrow}
\def\ri{\right}\def\ti{\tilde}\def\wti{\widetilde}
\def\al{\alpha}\def\bt{\beta}
\def\de{\delta}\def\De{\Delta}
\def\te{\theta}\def\la{\lambda}
\def\si{\sigma}\def\om{\omega}

\def\AB{{_{A,B}}}
\def\vec#1{{\bf #1}}
\begin{document}

%\twocolumn[\hsize\textwidth\columnwidth\hsize\csname@twocolumnfalse\endcsname  %\preprint{Imperial/TP/97-98/79, hep-th/98}

\title{Quantum Field Theory of boson mixing}

\author{Massimo Blasone${}^{\sharp \flat }$, Antonio
Capolupo${}^{\flat}$, Oreste Romei${}^{\flat}$
and Giuseppe Vitiello${}^{\flat }$ \vspace{3mm}}

 \address{${}^{\dag}$ Blackett Laboratory, Imperial College, Prince
Consort Road, London SW7 2BZ, U.K.
\\ [2mm] ${}^{\flat }$
Dipartimento di Fisica and INFN, Universit\`a di Salerno, I-84100
Salerno, Italy
\vspace{2mm}}

%\date{{\bf Version}, \today}

\maketitle

\begin{abstract}
We consider the quantum field theoretical formulation of boson field
mixing and obtain the exact oscillation formula. This formula
does not depend on arbitrary mass parameters. We show that
the space for the mixed field states is unitarily inequivalent to the
state space where the unmixed field operators are defined.
We also study the structure of the currents and charges
for the mixed fields.
\end{abstract}

\vspace{8mm}

%]

%\tableofcontents\newpage

%P.A.C.S.:

\section{Introduction}

Particle mixing and oscillations
(for a recent review, see Ref.\cite{zralek})
are among the most intriguing topics of
 Particle Physics. The mixing of neutrinos and their
oscillations seem to be now experimentally  established after a long
search \cite{kamiokande}.
On the other hand, quark mixing and meson mixing are
widely accepted and verified \cite{mesons}.
However, many features of the physics of mixing are still obscure, for
example the issue related to
its origin in the context of Standard Model
and the related problem of the generation of masses \cite{Fri}.
%Also the violation of
%CP symmetry is intimately connected with the structure of the
%Cabibbo--Kobayashi--Maskawa mixing matrix for quarks.

Also from a purely mathematical point of view, there are aspects
which are not fully understood. Indeed, only recently
\cite{BV95} a rich non--perturbative
vacuum structure has been discovered to be
associated with the mixing of fermion fields in the context of Quantum
Field Theory (for a mathematically rigorous approach see
Ref.\cite{hannabus}).
The careful study of such a structure \cite{Fujii:1999xa}
has led to the
determination of the exact QFT formula for neutrino
oscillations \cite{BHV98,Blasone:1999jb}, exhibiting
new features with respect to the usual quantum mechanical
Pontecorvo formula  \cite{BP78}.
Actually, it turns out \cite{lathuile,binger} that the non--trivial
nature of the mixing transformations manifests itself
also in the case of the
mixing of boson fields. Of course, in this case the condensate structure
for the ``flavor'' vacuum is very much different from the fermion case
and a careful analysis is necessary in order to understand which
phenomenological consequences are to be expected for the oscillations of
mixed bosons.

In this paper, we perform this analysis first at a formal level
and then we study the oscillations of mixed mesons of the kind of
the $K^0-{\bar K^0}$. We will treat these
particles as stable ones, an approximation which however does not
affect the general validity of our results. In the framework of
the QFT analysis of Refs. \cite{BV95,lathuile}, a study of the
meson mixing and oscillations has been carried out in
Ref.\cite{binger}, where modifications to the usual oscillation
formulas, connected with the vacuum structure, have been
presented. However, the results of Ref. \cite{binger} can be
improved in many respects and in the present paper we show that
the oscillation formula there obtained has to be actually replaced
with the exact one here presented.

In Section II we study the quantum field theory of two mixed
spin-zero boson fields.
In Section III we analyze the structure of currents for mixed fields
and we derive the exact oscillation formula in Section IV.
Section V is devoted to conclusions.
Some mathematical derivations are given in the Appendix.

%%%%%%%%%%%%%%%%%%%%%%%%%%%%%%%%%%%%%%%%%%%%%%%%%%%%%%%%%%%%%
\section{Mixing of boson fields in QFT}
%%%%%%%%%%%%%%%%%%%%%%%%%%%%%%%%%%%%%%%%%%%%%%%%%%%%%%%%%%%%%

The observed boson oscillations always involve particles with zero
electrical charge. What oscillate are some other quantum numbers such as
the strangeness and the isospin. Therefore, in the study of boson mixing
we have to consider always \cite{lurie} complex  fields.
The charge in question is some ``flavor charge'' (e.g. the strangeness)
and thus the complex fields are ``flavor charged'' fields, referred to
as ``flavor fields'' for simplicity.

We define the mixing relations as:
\bea\non
&&\phi_{A}(x) = \phi_{1}(x) \; \cos\te + \phi_{2}(x) \; \sin\te
\\[2mm] \lab{2.53}
&&\phi_{B}(x) =- \phi_{1}(x) \; \sin\te + \phi_{2}(x)\; \cos\te
\eea
where generically we denote the mixed fields with suffixes $A$ and $B$.
Let the fields $\phi_{i}(x)$, $i=1,2$, be free complex fields with
definite masses. Their
conjugate momenta are $\pi_{i}(x)=\pa_{0}\phi_{i}^{\dag}(x)$ and
the commutation relations are the usual ones:
\bea &&\lf[\phi_{i}(x),\pi_{j}(y)\ri]_{t=t'}=
\lf[\phi_{i}^{\dag}(x),\pi_{j}^{\dag}(y)\ri]_{t=t'}=i\de^{3}
(\vec{x}-\vec{y}) \, \de_{ij} \quad, \qquad i,j=1,2\,. \lab{2.50} \eea
with the other equal--time commutators vanishing.
The Fourier expansions of fields and momenta are:
\bea\lab{2.51} \phi_{i}(x) = \int \frac{d^3 k}{(2\pi)^{\frac{3}{2}}}
\frac{1}{\sqrt{2\om_{k,i}}} \lf( a_{{\bf k},i}\, e^{-i \om_{k,i} t} +
b^{\dag }_{{\bf k},i}\, e^{i \om_{k,i} t}  \ri) e^{i {\bf k}\cdot {\bf
x}} \eea
\bea\lab{2.52} \pi_{i}(x) = i\,\int \frac{d^3 k}{(2\pi)^{\frac{3}{2}}}
\sqrt{\frac{\om_{k,i}}{2}} \lf( a^{\dag }_{{\bf k},i}\, e^{i \om_{k,i}
t} - b_{{\bf k},i}\, e^{-i \om_{k,i} t} \ri) e^{i {\bf k}\cdot {\bf
x}}\,, \eea
where $\om_{k,i}=\sqrt{{\bf k}^2
+ m_i^2}$ and $\lf[a_{{\bf k},i},a_{{\bf p},j}^{\dag} \ri]= \lf[b_{{\bf
k},i},b_{{\bf p},j}^{\dag} \ri]=\de^{3}(\vec{k}-\vec{p}) \de_{ij}\, ,$
with $i,j=1,2\,$ and the other commutators vanishing. We will consider
stable particles, which will not affect the general validity of
our results.

We now proceed in a similar way to what has been done in
Ref.\cite{BV95} for fermions and recast Eqs.(\ref{2.53}) into the form:
\bea
\phi_{A}(x) = G^{-1}_\te(t)\; \phi_{1}(x)\; G_\te(t) \\[2mm]
\lab{2.53b}
\phi_{B}(x) = G^{-1}_\te(t)\; \phi_{2}(x)\; G_\te(t)
\eea
and similar ones for $\pi_{A}(x)$, $\pi_B(x)$. $G_\te(t)$  denotes the
operator which implements the mixing transformations (\ref{2.53}):
\bea\lab{2.54}
G_\te(t) = exp\lf[-i\;\te \int d^{3}x
\lf(\pi_{1}(x)\phi_{2}(x) - \phi_{1}^{\dag}(x)\pi_{2}^{\dag}(x)
-\pi_{2}(x)\phi_{1}(x) + \phi_{2}^{\dag}(x)\pi_{1}^{\dag}(x)\ri)\ri]\, ,
\eea
which is (at finite volume) a unitary operator:
$G^{-1}_\te(t)=G_{-\te}(t)=G^{\dag}_\te(t)$. The generator of the
mixing transformation in the exponent of $G_{\te}(t)$  can also be
written as
\bea\lab{2.55}
G_\te(t) = \exp[\te(S_{+}(t) - S_{-}(t))] ~.
\eea
The operators
\bea\lab{2.55b}
S_+(t)= S_{-}^{\dag}(t) \equiv -i\;\int d^3 x \;
(\pi_{1}(x)\phi_{2}(x) - \phi_{1}^{\dag}(x)\pi_{2}^{\dag}(x)) \,,
\eea
together with
\bea\lab{2.56}
S_{3} \equiv \frac{-i}{2} \int d^3 x
\lf(\pi_{1}(x)\phi_{1}(x) - \phi_{1}^{\dag}(x)\pi_{1}^{\dag}(x)
- \pi_{2}(x)\phi_{2}(x)
+\phi_{2}^{\dag}(x) \pi_{2}^{\dag}(x)
\ri)
\eea
\bea\lab{2.57}
S_{0} =\frac{Q}{2}\equiv \frac{-i}{2} \int d^3 x
\lf( \pi_{1}(x)\phi_{1}(x) -\phi_{1}^{\dag}(x)\pi_{1}^{\dag}(x)
+\pi_{2}(x)\phi_{2}(x) - \phi_{2}^{\dag}(x)\pi_{2}^{\dag}(x)
\ri)  ~,
\eea
close the $su(2)$ algebra (at each time $t$):
$[S_{+}(t) , S_{-}(t)]=2S_{3} $ , $ [S_{3} , S_{\pm}(t) ] = \pm
S_{\pm}(t)$ ,   $[S_{0} , S_{3}]= [S_{0} , S_{\pm}(t) ] = 0$.
Note that $S_{3}$ and $S_{0}$ are time independent.
It is useful to write down explicitly the expansions of the above
generators in terms of annihilation and creation operators:
\bea\lab{2.58}
S_{+}(t)=\int d^3 k \lf( U^*_{{\bf k}}(t) \, a_{{\bf
k},1}^{\dag}a_{{\bf k},2} -
V_{{\bf k}}^{*}(t) \, b_{-{\bf k},1}a_{{\bf k},2} + V_{{\bf k}}(t) \,
a_{{\bf k},1}^{\dag}b_{-{\bf k},2}^{\dag}
- U_{{\bf k}}(t) \, b_{-{\bf k},1}b_{-{\bf k},2}^{\dag} \ri)
\eea
\bea\lab{2.59}
S_{-}(t)=\int d^3 k \lf( U_{{\bf k}}(t) \, a_{{\bf
k},2}^{\dag}a_{{\bf k},1}
- V_{{\bf k}}(t) \, a_{{\bf k},2}^{\dag}b_{-{\bf k},1}^{\dag}
+ V_{{\bf k}}^{*}(t) \, b_{-{\bf k},2}a_{{\bf k},1}
- U^*_{{\bf k}}(t) \, b_{-{\bf k},2}b_{-{\bf k},1}^{\dag} \ri)
\eea
\bea\lab{2.58a}
S_{3}=\frac{1}{2}
\int d^3 k \lf( \, a_{{\bf k},1}^{\dag}a_{{\bf k},1} \,-
\, b_{-{\bf k},1}^\dag b_{-{\bf k},1}\, - \,
a_{{\bf k},2}^{\dag}a_{{\bf k},2}
\,+ \, b_{-{\bf k},2}^\dag b_{-{\bf k},2} \ri)
\eea
\bea\lab{2.58b}
S_{0}=\frac{1}{2}
\int d^3 k \lf( \, a_{{\bf k},1}^{\dag}a_{{\bf k},1} \,-
\, b_{-{\bf k},1}^\dag b_{-{\bf k},1}\, + \,
a_{{\bf k},2}^{\dag}a_{{\bf k},2}
\,- \, b_{-{\bf k},2}^\dag b_{-{\bf k},2} \ri)\,.
\eea

As for the case of the fermion mixing, the structure of  the generator
Eq.(\ref{2.54}) is recognized to be the one of a rotation
combined with a
Bogoliubov transformation (see below Eqs.(\ref{2.62a})-(\ref{2.62d})).
Indeed, in the above equations,  the
coefficients $U_{{\bf k}}(t)\equiv |U_{{\bf k}}| \; e^{i(\om_{k,2}-
\om_{k,1})t}$ and $V_{{\bf k}}(t)\equiv |V_{{\bf k}}| \;
e^{i(\om_{k,1}+ \om_{k,2})t}$ appear to be the Bogoliubov coefficients.
They are defined as
\bea
&&|U_{{\bf k}}|\equiv \frac{1}{2}
\lf( \sqrt{\frac{\om_{k,1}}{\om_{k,2}}}
+ \sqrt{\frac{\om_{k,2}}{\om_{k,1}}}\ri) ~,
\;\;\;\;\;\;
|V_{{\bf k}}|\equiv  \frac{1}{2} \lf(
\sqrt{\frac{\om_{k,1}}{\om_{k,2}}}
- \sqrt{\frac{\om_{k,2}}{\om_{k,1}}} \ri)  \eea
and satisfy the relation
\bea\lab{2.60}
&&|U_{{\bf k}}|^{2}-|V_{{\bf k}}|^{2}=1\,,
\eea
which is in fact to be expected in  the boson case
(note the difference
with respect to the fermion case
of Ref.\cite{BV95}).  We can thus  put $|U_{{\bf k}}|\equiv \cosh
\xi^{\bf k}_{1,2}$ , $|V_{{\bf k}}|\equiv \sinh \xi^{\bf k}_{1,2}$,
with
$\xi^{\bf k}_{1,2}= \frac{1}{2} \ln\frac{\om_{k,1}}{\om_{k,2}}$.

We now consider the action of the generator of the
mixing transformations on the vacuum $|0 \ran_{1,2}$ for the
fields $\phi_{1,2}(x)$:  $a_{{\bf k},i}|0 \ran_{1,2} = 0, ~ i=1,2$ .
The generator induces an $SU(2)$ coherent state structure on such
state \cite{Per}:
\bea\label{2.61} |0(\te, t) \ran_\AB \equiv G^{-1}_\te(t)\; |0
\ran_{1,2}\,. \eea
From now on we will refer to the state $|0(\te, t) \ran_\AB$ as to
the ``flavor'' vacuum for bosons. The suffixes $A$ and $B$ label
the flavor charge content of the state. We have
$\,_\AB\lan 0(\te,t)|0 (\te,t)\ran_\AB \,= \,1$.
In the following, we will
consider the Hilbert space for flavor fields at a given time $t$,
say $t=0$, and it is useful to define $|0 (t)\ran_\AB\equiv|0(\te,
t)\ran_\AB$ and $|0 \ran_\AB\equiv|0(\te, t=0)\ran_\AB$ for future
reference. A crucial point is that the flavor and the mass vacua
are orthogonal in the infinite volume  limit. We indeed have (see
Appendix):
\bea\label{2.61a} \,_{1,2}\lan 0|0 (t)\ran_\AB \,= \,
\prod\limits_{\bf k} \, _{1,2}\lan 0| G_{{\bf k},\te}^{-1}(t)
|0\ran_{1,2}\,=\, \prod\limits_{\bf k} \,f_0^{\bf k}(\te) ~,
~~~for ~ any ~ t, \eea
where we have used $G^{-1}_\te(t) = \prod\limits_{\bf k} \, G_{{\bf
k},\te}^{-1}(t)$ (see Eqs.(\ref{2.55}), (\ref{2.58}) and (\ref{2.59})
). In the infinite volume limit, we obtain
\bea\label{2.61b}
\lim\limits_{V\rar \infty}\,_{1,2}\lan 0|0 (t)\ran_\AB
= \lim\limits_{V\rar \infty}\, e^{\frac{V}{(2\pi)^3}
\int d^3 k \, \ln
f_0^{\bf k}(\te) } \, = \, 0 ~, ~~~for ~ any ~ t. \eea
From the Appendix, Eq.(\ref{A74}), we see that $\ln f_0^{\bf
k}(\te)$ is indeed negative for any values of ${\bf k}$, $\te$ and
$m_1, m_2$ (note that $0\le \te \le \pi/4$). We also observe that
the orthogonality disappears when   $\te=0$ and/or $m_1=m_2$,
consistently with the fact that in both cases  there is no mixing.
These features are similar to the case of fermion  mixing
\cite{BV95}: the orthogonality is essentially due to the infinite
number of degrees of freedom \cite{Itz,Um1}
(The statement of Ref.\cite{binger}
that  in the boson case the above vacua are orthogonal also at
finite volume has to be therefore corrected according to the
present result).
%It is important to note that the orthogonality is
%also between the states $|0 (\te, t)\ran_\AB $ and $|0 (\te',
%t')\ran_\AB$, for $\te \neq \te'$ and/or $t\neq t'$
%(see Appendix B).

%
We can define annihilation operators  for the vacuum $|0(t) \ran_\AB$
as $a_{{\bf k},A}(\te ,t) \equiv G^{-1}_\te(t) \; a_{{\bf
k},1}\;G_\te(t)$, etc.. with $a_{{\bf k},A}(\te ,t) |0(t) \ran_\AB =
0$. For simplicity we will use the notation $ a_{{\bf k},A}(t) \equiv
a_{{\bf k},A}(\te ,t)$. Explicitly, we have:
\bea \label{2.62a}
a_{{\bf k},A}(t)&=&\cos\te\;a_{{\bf k},1}\;+\;\sin\te\;\lf(
U^*_{{\bf k}}(t)\; a_{{\bf k},2}\;+\; V_{{\bf k}}(t)\;
b^{\dag}_{-{\bf k},2}\ri)\, ,
\\
a_{{\bf k},B}(t)&=&\cos\te\;a_{{\bf k},2}\;-\;\sin\te\;\lf(
U_{{\bf k}}(t)\; a_{{\bf k},1}\;-
\; V_{{\bf k}}(t)\; b^{\dag}_{-{\bf k},1}\ri)\, ,
\\
b_{-{\bf k},A}(t)&=&\cos\te\;b_{-{\bf k},1}\;+\;\sin\te\;\lf(
U^*_{{\bf k}}(t)\; b_{-{\bf k},2}\;+
\; V_{{\bf k}}(t)\; a^{\dag}_{{\bf k},2}\ri)\, ,
\\ \label{2.62d}
b_{-{\bf k},B}(t)&=&\cos\te\;b_{-{\bf k},2}\;-\;\sin\te\;\lf(
U_{{\bf k}}(t)\; b_{-{\bf k},1}\;-
\; V_{{\bf k}}(t)\; a^{\dag}_{{\bf k},1}\ri) ~.
\eea
These operators satisfy the canonical commutation relations (at equal
times). In their expressions the Bogoliubov transformation part is
evidently characterized by the terms with the $U$ and $V$ coefficients.
The condensation density of the flavor vacuum is given for any $t$ by
\bea\label{2.63} {}_\AB\lan 0(t)| a_{{\bf k},i}^{\dag} a_{{\bf k},i}
|0(t)\ran_\AB= {}_\AB\lan 0(t)| b_{-{\bf k},i}^{\dag} b_{-{\bf k},i}
|0(t)\ran_\AB=\sin^{2}\te\;  |V_{{\bf k}}|^{2} \quad, \qquad i=1,2\,.
\eea

It is useful to note that $|V_{{\bf k}}|^{2}$ can be written as a
function of the rescaled momentum
$p\equiv\sqrt{\frac{2 |{\bf k}|^2}{m_1^2
+m_2^2}}$ and of the adimensional parameter $a\equiv \frac{m_2^2
-m_1^2}{m_1^2 +m_2^2}$ as follows:
\bea
|V(p,a)|^2 & =& \frac{p^2 +1}{2\sqrt{(p^2 + 1)^2 - a^2}}
-\frac{1}{2} ~,
\eea
from which we see that the condensation density is maximal at $p=0$
($|V_{max}|^2 = \frac{(m_1 -m_2)^2}{4 m_1 m_2}$) and goes
to zero for large momenta (i.e. for $|{\bf k}|^2\gg
\frac{m_1^2 +m_2^2}{2}$). Note that the corresponding quantity in the
fermion case is limited to the value $1/2$ and the momentum scale is
given by $\sqrt{m_1 m_2}$. A plot of $|V(p,a)|^2$
is presented in Fig.1 for sample values of the parameter $a$.

\vspace{.5cm}
\centerline{\epsfysize=3.0truein\epsfbox{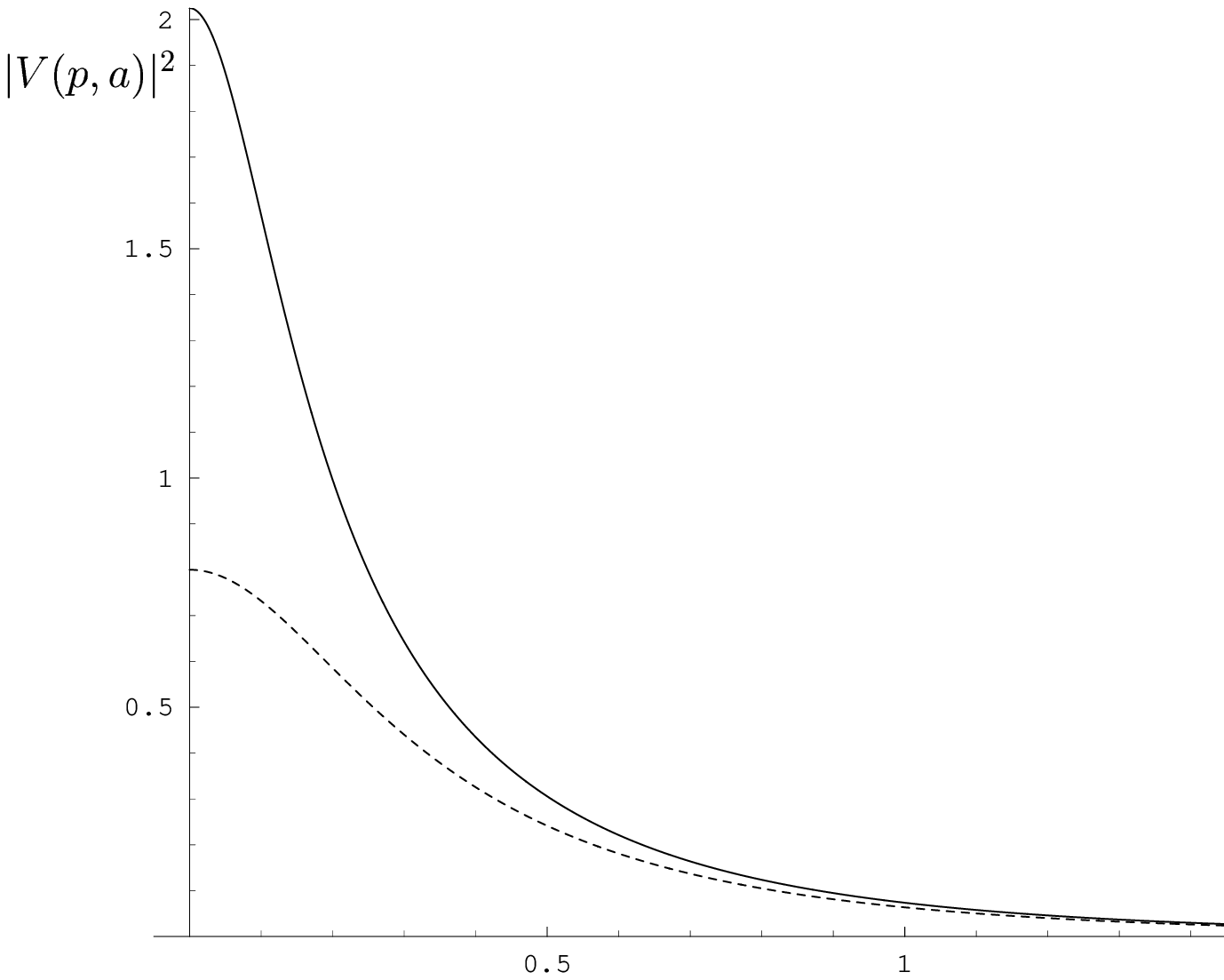}}
\vspace{.2cm}
\centerline{\small Figure 1: The condensation density
$|V(p,a)|^2$ as a function of $p$
for $a=0.98$ (solid line) and $a=0.92$ (dashed line). }

%%%%%%%%%%%%%%%%%%%%%%%%%%%
\subsection{Arbitrary mass parameterization}
%%%%%%%%%%%%%%%%%%%%%%%%%%%

Above we have expanded the mixed fields $\phi_{A,B}$ in the same
basis as the free fields $\phi_{1,2}$. However, as noticed in
Ref.\cite{Fujii:1999xa} for the case of fermion mixing,
this is not the most general possibility. Indeed, one could as
well expand the flavor fields in a basis of fields with
arbitrary masses. Of course, these arbitrary mass parameters
should not appear in the physically observable quantities. Thus,
as a check for the validity of the oscillation formula we are
going to derive in Section IV, it is  important to consider this
generalization. Let us first rewrite the free fields $\phi_{1,2}$
in the form
\bea\label{gener1}
\phi_{i}(x) &=& \int
\frac{d^3 k}{(2\pi)^{\frac{3}{2}}}
\lf(u^\phi_{{\bf k},i}(t)\, a_{{\bf k},i}\, +
\, v^\phi_{-{\bf k},i}(t)
b^{\dag }_{-{\bf k},i} \ri)
e^{i {\bf k}\cdot {\bf x}} \, ,
\\[3mm] \label{gener2}
\pi_{i}(x) & =& i\,\int
\frac{d^3 k}{(2\pi)^{\frac{3}{2}}}
\lf( u^\pi_{{\bf k},i}(t)\,a^{\dag }_{{\bf k},i}\: -
\,v^\pi_{-{\bf k},i}(t)\, b_{-{\bf k},i}
\ri)e^{i {\bf k}\cdot {\bf x}} \qquad , \qquad i=1,2,
\eea
where we have introduced the notation
\bea
&&u^\phi_{{\bf k},i}(t)\, \equiv \,\frac{1}{\sqrt{2 \om_{k,i}}} e^{-i
\om_{k,i}t} \quad ,\quad v^\phi_{-{\bf k},i}(t)\,\equiv \,
\frac{1}{\sqrt{2 \om_{k,i}}}e^{i \om_{k,i}t} ~,
\\ [2mm]
&&u^\pi_{{\bf k},i}(t)\,\equiv\,\sqrt{\frac{\om_{k,i}}{2}}
e^{i \om_{k,i}t}\qquad, \quad v^\pi_{-{\bf k},i}(t)\,\equiv \,
\sqrt{\frac{\om_{k,i}}{2}} e^{-i \om_{k,i}t} \qquad , \qquad i=1,2~.
\eea

We now define
\bea\label{rho}
\rho^{\bf k *}_{\al\bt}(t)& \equiv& u^\pi_{{\bf k},\al}(t)
u^\phi_{{\bf k},\bt}(t) \,+\, v^\phi_{-{\bf k},\al}(t)v^\pi_{-{\bf
k},\bt}(t) \,=\,
%\frac{1}{2}\lf[\sqrt{\frac{\om_{k,\al}}{\om_{k,\bt}}}
%+\sqrt{\frac{\om_{k,\bt}}{\om_{k,\al}}} \,\ri]
e^{i (\om_{k,\al} - \om_{k,\bt}) t}\,\cosh\,\xi_{\al,\bt}^{\bf k} ~,
\\ [2mm]
\la^{\bf k *}_{\al\bt}(t)& \equiv& v^\pi_{-{\bf k},\al}(t)
u^\phi_{{\bf k},\bt}(t) \,-\, u^\phi_{{\bf k},\al}(t)v^\pi_{-{\bf
k},\bt}(t)\,= \,
%\frac{1}{2}\lf[\sqrt{\frac{\om_{k,\al}}{\om_{k,\bt}}}
%-\sqrt{\frac{\om_{k,\bt}}{\om_{k,\al}}} \,\ri]
e^{-i (\om_{k,\al} + \om_{k,\bt}) t} \,
\sinh\,\xi_{\al,\bt}^{\bf k}~,
\\ [2mm]
\xi_{\al,\bt}^{\bf k}&\equiv & \frac{1}{2}
\ln\frac{\om_{k,\al}}{\om_{k,\bt}} \qquad , \qquad \al,\bt = 1,2,A,B ,
\eea
where $\om_{k,\al}\equiv\sqrt{k^2 +
\mu_\al^2}$. We denote with  $\mu_A$ and $\mu_B$ the
arbitrary mass parameters while $\mu_1\equiv m_1$
and $\mu_2\equiv m_2$ are the  physical masses.
Note that $\rho^{\bf k }_{1 2}(t)=U_{\bf
k}(t)$ and $\la^{\bf k }_{1 2}(t)=V_{\bf k }(t)$.
We can now write the expansion of the
flavor fields in the general form (we use a tilde to
denote the generalized ladder operators):
\bea\label{gener7}
\phi_{\si}(x) = \int
\frac{d^3 k}{(2\pi)^{\frac{3}{2}}}
\lf(u^\phi_{{\bf k},\si}(t)\, {\ti a}_{{\bf k},\si}(t) \,
+ \, v^\phi_{-{\bf k},\si}(t) \,
{\ti b}^{\dag }_{-{\bf k},\si}(t) \ri)
e^{i {\bf k}\cdot {\bf x}} \qquad , \qquad  \si = A,B~,
\eea
which is to be compared with the expansion in the free field basis as
given in Eqs. (\ref{2.53b}):
\bea\label{gener8}
\phi_{\si}(x) = \int \frac{d^3 k}{(2\pi)^{\frac{3}{2}}}
\lf(u^\phi_{{\bf k},i}(t)\, a_{{\bf k},\si}(t) \,
+ \, v^\phi_{-{\bf k},i}(t) \,b^{\dag }_{-{\bf k},\si}(t) \ri)
e^{i {\bf k}\cdot {\bf x}}~,
\eea
where $(\si,i)=(A,1),(B,2)$.
The relation between the two sets of flavor operators is given as
\bea \label{bogol}
\lf(\ba{c} {\ti a}_{{\bf k},\si}(t)
\\ [2mm] {\ti b}^{\dag }_{-{\bf  k},\si}(t) \ea\ri) & =& J^{-1}(t)
\lf(\ba{c} a_{{\bf k},\si}(t)
\\ [2mm] b^{\dag }_{-{\bf k},\si}(t)\ea\ri)J(t)
\, =\,\lf(\ba{cc} \rho^{\bf k *}_{\si i}(t) &
\la^{\bf k}_{\si i}(t)
\\ [2mm]\la^{\bf k *}_{\si i}(t) &
\rho^{\bf k}_{\si i}(t)  \ea\ri) \,\lf(\ba{c}
a_{{\bf k},\si}(t)
\\ [2mm]
b^{\dag }_{-{\bf k},\si}(t) \ea\ri) ~,
\\ [2mm]
J(t)&=& \exp\left\{ \int d^3 k \, \xi_{\sigma,i}^{\bf k}\left[
a^{\dag}_{{\bf k},\sigma}(t)b^{\dag}_{{-\bf k},\sigma}(t) -
b_{{-\bf k},\sigma}(t)a_{{\bf k},\sigma}(t)  \right]\right\}\, ,
\eea
with $\xi_{\sigma,i}^{\bf k}\equiv \frac{1}{2}
\ln\frac{\om_{k,\si}}{\om_{k,i}}$. For $\mu_A=m_1$ and $\mu_B=m_2$
one has $J = 1$. Note that the transformation Eq.(\ref{bogol}) is
in fact a Bogoliubov transformation which leaves invariant the
form $a^{\dag}_{{\bf k},\sigma}(t)a_{{\bf k},\sigma}(t) -
b^{\dag}_{{-\bf k},\sigma}(t)b_{{-\bf k},\sigma}(t)$.

%%%%%%%%%%%%%%%%%%%%%%%%%%%%%%%%%%%%%%%%%%%%%%%%%%%%%%%%%%%%
\section{The currents for mixed boson fields}
%%%%%%%%%%%%%%%%%%%%%%%%%%%%%%%%%%%%%%%%%%%%%%%%%%%%%%%%%%%%

Before presenting the exact oscillation formula, let us investigate in
this Section the structure of currents and charges for the mixed
fields. This will enable us to identify the relevant physical
observables to look at for flavor oscillations. Since we are here
interested in vacuum oscillations, in the following we neglect
interaction terms and only consider the free field Lagrangian for
two charged scalar fields with a mixed mass term:
\bea\label{boslagAB}
{\cal L}(x)&=& \pa_\mu \Phi_f^\dag(x)\, \pa^\mu \Phi_f (x)\, - \,
\Phi_f^\dag(x)  M \Phi_f(x) \, ,
\eea
with $\Phi_f^T=(\phi_A,\phi_B)$, $ M =
\lf(\ba{cc} m_A^2 & m_{AB}^2 \\ m_{AB}^2 & m_B^2\ea \ri)$. By means of
Eq.(\ref{2.53}),
\bea\label{bosmix}
\Phi_f(x) \, =\, \lf(\ba{cc} \cos \te & \sin \te \\[2mm]
-\sin \te & \cos \te \ea\ri) \Phi_m (x) ~,
\eea
${\cal L}$ becomes diagonal in the basis
$\Phi_m^T=(\phi_1,\phi_2)$:
\bea \label{boslag12}
{\cal L}(x)&=&\pa_\mu \Phi_m^\dag(x) \,\pa^\mu \Phi_m(x) \, - \,
\Phi_m^\dag(x)  M_d \Phi_m (x) ~,
\eea
where $M_d =
diag(m_1^2,m_2^2)$ and
$ m_A^2 = m_1^2\cos^{2}\te +
m_2^2 \sin^{2}\te~$,  $m_B^2 = m_1^2\sin^{2}\te + m_2^2
\cos^{2}\te~$,   $m_{AB}^2 =(m_2^2-m_1^2)\sin\te \cos\te\,$.

The Lagrangian  ${\cal L}$ is invariant under the global $U(1)$
phase transformations  $\Phi_m' \, =\, e^{i \al }\, \Phi_m$: as a
result, we have the conservation of the Noether  charge $Q=\int d^3x \,
I^0(x)$, which is indeed the total charge of the system (we have
$I^\mu(x)= i\, \Phi_m^\dag(x) \, \stackrel{\lrar}{\pa^\mu}\, \Phi_m (x)
$ with $ \stackrel{\lrar}{\pa^\mu}  \,\equiv \,\stackrel{\rar}{\pa^\mu}
- \stackrel{\lar}{\pa^\mu}   $).

Let us now consider the $SU(2)$ transformation
\bea \label{masssu2}
\Phi_m'(x) \, =\, e^{i \al_j \tau_j}\, \Phi_m(x) \qquad,
\qquad j= 1,2,3 \, ,
\eea
with $\al_j$ real constants, $\tau_j=\si_j/2$ and $\sigma_j$ being
the Pauli matrices. For $m_1\neq m_2$,
the Lagrangian is not generally invariant under
(\ref{masssu2}) and we obtain,
by use of the equations of motion,
\bea \label{boscu1}
\de {\cal L}(x)&= &  - i \,\al_j\, \Phi_m^\dag(x)
\, [M_d\,, \,\tau_j] \, \Phi_m(x) \, = \, -\al_j \,\pa_\mu\,
J^\mu_{m,j}(x)\, ,
\\ [2mm]
J^\mu_{m,j}(x) &=& i\, \Phi_m^\dag(x) \, \tau_j\,
\stackrel{\lrar}{\pa^\mu}\, \Phi_m (x) \quad , \qquad j= 1,2,3  .
\eea
The  corresponding charges, $Q_{m,j}(t)\equiv \int d^3 x
\,J^0_{m,j}(x) $, close  the $su(2)$ algebra (at each time $t$). The
Casimir operator $C_m$ is proportional to the total charge:
$C_m \equiv
\Big[ \sum\limits_{j=1}^3 Q_{m,j}^2(t)\Big]^{\frac{1}{2}}$
$=\,\frac{1}{2}Q$. Observe also that
the transformation induced by $Q_{m,2}(t)$,
\bea
\Phi_f(x) &=& e^{-2 i\te Q_{m,2}(t)} \Phi_m(x) e^{2i\te Q_{m,2}(t)}
\eea
is just the mixing transformation Eq.(\ref{bosmix}).
Thus $ 2 Q_{m,2}(t)$ is the
generator of the mixing transformations. Moreover, $
Q_{m, \pm}(t) \equiv {\frac{1}{2}}(Q_{m,1}(t) \pm iQ_{m,2}(t))$,
$Q_{m,3}$, and $C_{m}$ are nothing but $S_\pm(t)$, $S_{3}$, and
$S_{0}$, respectively, as introduced in Eqs.(\ref{2.55b})-(\ref{2.57}).
From Eq.(\ref{boscu1}) we also see  that $Q_{m,3}$ and $C_{m}$ are
conserved, consistently with Eqs.(\ref{2.58a})-(\ref{2.58b}). Observe
that the combinations
\bea
Q_{1,2}&\equiv& \frac{1}{2}Q \pm Q_{m,3}
\\[2mm] \label{noether1}
Q_i & = & \int d^3 k  \lf( a^{\dag}_{{\bf k},i} a_{{\bf k},i}\,
-\, b^{\dag}_{-{\bf k},i}b_{-{\bf k},i}\ri) \quad ,\qquad i=1,2,
\eea
are simply the conserved\footnote{Note that, in absence of
mixing, these charges would indeed be the flavor charges, being the
flavor conserved for each generation.} (Noether) charges for the free
fields $\phi_1$ and $\phi_2$ with $Q_1 + Q_2 = Q$.

We now perform the $SU(2)$ transformations on the flavor doublet
$\Phi_f$:
\bea \Phi_f'(x) \, =\, e^{ i  \al_j \tau_j }\,
\Phi_f (x) \quad ,\qquad
j \,=\, 1, 2, 3,
\eea
and obtain:
\bea \de {\cal L}(x)&= &
-i \,\al_j\,\Phi_f^\dag(x)\, [M,\tau_j]\,
\Phi_f(x) \, = \, -\al_j \,\pa_\mu  J_{f,j}^{\mu}(x) ~,
\\ [3mm]
J^\mu_{f,j}(x) &=&  i \,\Phi_f^\dag(x) \, \tau_j\,
\stackrel{\lrar}{\pa^\mu}\, \Phi_f (x) \quad ,\qquad j \,=\, 1, 2, 3.
\eea

The related charges, $Q_{f,j}(t)$ $\equiv$  $\int d^3 x \,J^0_{f,j}(x)
$, still fulfil the $su(2)$ algebra and $C_f=C_m =\frac{1}{2}Q$.
Due to
the off--diagonal (mixing) terms in the mass matrix $M$, $Q_{f,3}(t)$
is time--dependent. This implies an exchange of charge between $\phi_A$
and $\phi_B$, resulting in the flavor oscillations.
This suggests to us
to define indeed the {\em flavor charges} as
\bea \label{flacha}
Q_A(t) & \equiv &\frac{1}{2}Q  \, + \, Q_{f,3}(t)~,
\\[2mm]
Q_B(t) & \equiv & \frac{1}{2}Q \, -  \, Q_{f,3}(t)~,
\eea
with $Q_A(t) \, + \,Q_B(t) \, = \, Q$. These charges have a simple
expression in terms of the flavor  ladder operators:
\bea\label{flacha2}
 Q_\si(t) & = & \int d^3 k  \lf( a^{\dag}_{{\bf k},\si}(t)
a_{{\bf k},\si}(t)\,
-\, b^{\dag}_{-{\bf k},\si}(t)b_{-{\bf k},\si}(t)\ri)
\quad ,\qquad\si= A, B ~.
\eea

This is because they are connected to the Noether charges $Q_i$ of
Eq.(\ref{noether1}) via the mixing generator: $Q_\si(t) =
G^{-1}_{\te}(t)Q_i G_{\te}(t)$, with $(\si,i)=(A,1),(B,2)$.
Note that the flavor charges are invariant under the transformation
Eq.(\ref{bogol}).

%%%%%%%%%%%%%%%%%%%%%%%%%%%%%%%%%%%%%%%%%%%%%%%%%%%%%%%%%%%
\section{The oscillation formula for mixed bosons}
%%%%%%%%%%%%%%%%%%%%%%%%%%%%%%%%%%%%%%%%%%%%%%%%%%%%%%%%%%%

Let us now calculate the oscillation formula for mixed bosons. We will
first follow the approach of Ref. \cite{binger} and show that the
oscillation formulas there presented  exhibit a dependence on the
arbitrary mass parameters $\mu_\si$, a feature which is not physically
acceptable. We will do this by using the generalized operators
introduced above. Then  we will show how to cure this pathology, in
analogy to what was done in Refs. \cite{BHV98,Blasone:1999jb}, where
the exact formula for neutrino oscillations was derived and it was
shown to be independent from the arbitrary mass parameters that can be
introduced in the expansions of the flavor  fields.

\subsection{The oscillation formula of Binger and Ji}

Following Ref.\cite{binger}, let us define the (generalized) flavor
state by acting on the {\em mass vacuum} $|0\ran_{1,2}$ with the flavor
creation operators (we omit momentum indices):
\bea\label{BJstate}
 |{\wti a}_A \ran \equiv {\wti a}_A^\dag |0\ran_{1,2} \,=\,
\rho_{A 1} \cos\te |a_1 \ran \, + \, \rho_{A 2} \sin\te |a_2 \ran   ~.
\eea
with $|a_i \ran =a^\dag_i |0\ran $.
As already discussed in Ref.\cite{binger}, the flavor state so defined
is not normalized and the normalization factor has to be introduced as
\bea\label{normaliz}
{\wti {\cal N}}_A \equiv
\lan {\wti a}_A | {\wti a}_A \ran \,=\, \rho^2_{A 1} \cos^2\te +
\rho^2_{A 2} \sin^2\te  ~.
\eea
We have
\bea \label{na}
\lan {\wti a}_A | {\wti N}_A |{\wti a}_A \ran &=&( \rho^2_{A 1} +
\la^2_{A 1} ) \cos^2 \te + ( \rho^2_{A 2} + \la^2_{A 2} ) \sin^2 \te  ~.
\eea

The oscillation formula then follow as:
\bea \label{nat}
\lan {\wti a}_A(t) | {\wti N}_A |{\wti a}_A(t) \ran
&=& \lan {\wti a}_A | {\wti N}_A |{\wti a}_A \ran
- 4 \frac{ \rho^2_{A 1} \rho^2_{A 2} }{ {\wti {\cal N}}_A }
\sin^2 \te \cos^2 \te \sin^2 \lf( \frac{\De E}{2} t \ri)
\eea
and a similar one for the expectation value of ${\wti N}_B$.
From the above,
as announced, it is evident that these formulas explicitly depend on
the (arbitrary) parameters $\mu_\si$ (see Eq.(\ref{rho})).
We also note
that, for $\mu_A=m_1$ and $\mu_B=m_2$, one has $\rho_{A 1}=1$,
$\la_{A 1} =0 $, $\rho_{A 2}=U$ and $\la_{A 2} =V$. Consequently,
Eqs.(\ref{na}), (\ref{nat}) reduce respectively
to Eqs.(18) and (20) of Ref.\cite{binger}.

\subsection{The exact oscillation formula}

We now show how a consistent treatment of the flavor oscillation for
bosons in QFT can be given which does not exhibit the above
pathological dependence on arbitrary parameters.

There are two key points to be remarked. A general feature of field
mixing is  that the number operator  for mixed particles is not a
well-defined operator. It is so because the mixing transformations mix
creation and annihilation operators and then the annihilation
(creation) operators for flavor particles  and antiparticles do not
commute at different times (see Eqs.(\ref{2.62a})-(\ref{2.62d})).
Moreover, the number operator does depend on the arbitrary mass
parameters. Much care is therefore required in the use of the  number
operator. A second remark is that the flavor states are not to be
defined by using the vacuum $|0\ran_{1,2}$: the flavor states so
defined are in fact not normalized and the normalization factor
Eq.(\ref{normaliz}) depends on the arbitrary mass parameters.

These two difficulties can be bypassed by using the remedy already
adopted  in Refs.\cite{BHV98,Blasone:1999jb} for the case of fermions:
the flavor states have shown to be consistently defined by acting with
the flavor creation operators on the {\em flavor vacuum}. The
observable quantities are then the expectation values of the {\em
flavor charges} on the flavor states: the oscillation formulas thus
obtained do not depend on the arbitrary mass parameters.

In the line of Refs.\cite{BHV98,Blasone:1999jb}, let us now
define\footnote{In the following, we will work in the Heisenberg
picture: this is particularly convenient in the present
context since special care has to be taken with the
time dependence of flavor states
(see the discussion in Ref.\cite{BHV98}).} the state of
the $a_A$ particle as $|{\ti a}_{{\bf k},A}\ran_\AB\equiv$ $ {\ti
a}_{{\bf k},A}^\dag(0) |0\ran_\AB$ and consider the expectation
values of the flavor charges (\ref{flacha2}) on it (analogous
results follow if one considers $|{\ti a}_{{\bf k},B}\ran_\AB$).
We obtain:
\bea\label{sicharges}
 {\wti {\cal Q}}_{{\bf k},\si}(t)&\equiv& {}_\AB\lan {\wti
a}_{{\bf k},A} | {\wti Q}_\si(t) | {\wti a}_{{\bf k},A}\ran_\AB \,
=\, \lf|\lf[{\wti a}_{{\bf k},\si}(t), {\wti a}^{\dag}_{{\bf
k},A}(0) \ri]\ri|^2 \; - \; \lf|\lf[{\wti b}^\dag_{-{\bf
k},\si}(t), {\wti a}^{\dag}_{{\bf k},A}(0) \ri]\ri|^2 \quad,
\qquad \si=A,B\, . \eea
We also have $\,_\AB\langle 0| {\wti Q}_{{\bf k},\si}(t)| 0\rangle_\AB
=0$ and ${\wti {\cal Q}}_{{\bf k},A}(t) + {\wti {\cal Q}}_{{\bf
k},A}(t)= 1$.

A straightforward direct calculation shows that the above
quantities {\em do not depend} on $\mu_A$ and $\mu_B$, i.e.:
\bea \label{independence} {}_\AB\lan {\wti a}_{{\bf k},A} | {\wti
Q}_{{\bf k},\si}(t) | {\wti a}_{{\bf k},A}\ran_\AB \, =\,{}_\AB\lan
a_{{\bf k},A} | Q_{{\bf k},\si}(t) | a_{{\bf k},A}\ran_\AB \quad,
\qquad \si=A,B\, , \eea
and similar one for the expectation values on $|{\ti a}_{{\bf
k},B}\ran_\AB$. Eq.(\ref{independence}) is a central result of our
work: it confirms that the only physically relevant quantities are the
above expectation values of flavor charges. Note that expectation
values of the number operator, of the kind ${}_\AB\lan {\wti a}_{{\bf
k},A} | {\wti N}_\si(t) | {\wti a}_{{\bf k},A}\ran_\AB \, =\,
\lf|\lf[{\wti a}_{{\bf k},\si}(t), {\wti a}^{\dag}_{{\bf k},A}(0)
\ri]\ri|^2$ and similar ones, do indeed depend on the arbitrary mass
parameters, although the flavor states are properly defined (i.e. on
the flavor Hilbert space). The cancellation of these parameters happens
{\em only} when considering the combination of squared modula of
commutators of the form Eq.(\ref{sicharges})\footnote{One may think it
could make sense to take the expectation value of the flavor charges on
states defined on the mass Hilbert space, as the ones defined in
Eq.(\ref{BJstate}). A direct calculation however shows that this is not
the case and these expectation values depends on the mass parameters:
the conclusion is that one must use the flavor Hilbert space.}.
 A similar cancellation
occurs for fermions \cite{Blasone:1999jb} with the sum of the squared
modula of anticommutators.

Finally, the explicit calculation gives
\bea  \non
 {\cal Q}_{{\bf k},A}(t)&=&
\lf|\lf[a_{{\bf k},A}(t), a^{\dag}_{{\bf k},A}(0) \ri]\ri|^2 \; - \;
\lf|\lf[b^\dag_{-{\bf k},A}(t), a^{\dag}_{{\bf k},A}(0) \ri]\ri|^2
\\[2mm] \label{Acharge}
&=& 1 - \sin^{2}( 2 \theta) \lf[ |U_{{\bf k}}|^{2} \;
\sin^{2} \lf( \frac{\omega_{k,2} - \omega_{k,1}}{2} t \ri)
-|V_{{\bf k}}|^{2} \;
\sin^{2} \lf( \frac{\omega_{k,2} + \omega_{k,1}}{2} t \ri) \ri]
\, ,
\\[4mm]  \non
{\cal Q}_{{\bf k},B}(t)&=& \lf|\lf[a_{{\bf k},B}(t),
a^{\dag}_{{\bf k},A}(0) \ri]\ri|^2 \; - \; \lf|\lf[b^\dag_{-{\bf
k},B}(t), a^{\dag}_{{\bf k},A}(0) \ri]\ri|^2
\\[2mm] \label{Bcharge}
&=& \sin^{2}( 2 \theta)\lf[ |U_{{\bf k}}|^{2} \; \sin^{2} \lf(
\frac{\omega_{k,2} - \omega_{k,1}}{2} t \ri) -|V_{{\bf k}}|^{2} \;
\sin^{2} \lf( \frac{\omega_{k,2} + \omega_{k,1}}{2} t \ri) \ri] \, .
\eea

Notice the negative sign in front of the $|V_{{\bf k}}|^{2}$ terms
in these formulas, in contrast with the fermion case
\cite{BHV98,Blasone:1999jb}: the boson
flavor charge can assume also negative values. This fact  points
to the statistical nature of the phenomenon: it means that when
dealing with mixed fields, one intrinsically deals with a
many--particle system, i.e. a genuine field theory phenomenon.
This situation has a strong analogy with Thermal Field Theory
(i.e. QFT at finite temperature) \cite{Um1}, where quasi--particle
states are ill defined and only statistical averages make sense.
Of course, there is no violation of charge conservation for the
overall system of two mixed fields.

The above formulas are obviously different from the usual quantum
mechanical oscillation formulas, which however are recovered in the
relativistic limit (i.e. for $|{\bf k}|^2\gg
\frac{m_1^2 +m_2^2}{2}$). Apart
from the extra oscillating term (the one proportional to $|V_{{\bf
k}}|^{2}$) and the momentum dependent amplitudes, the QFT formulas
carry the remarkable information about the statistics of the
oscillating particles: for bosons and fermions the amplitudes
(Bogoliubov coefficients) are drastically different according to the
two different statistics ($|U_{\bf k}|$ and $|V_{\bf k}|$ are
circular functions in the fermion case and hyperbolic functions
in the boson case).
This fact also fits with the above mentioned
statistical nature of the oscillation phenomenon in QFT. Note also that our
treatment is essentially
 non--perturbative and in this differs from other QFT approaches to
particle mixing and oscillations
(see for example Ref.\cite{Beuthe:2000er} for a review).

In order to better appreciate the features of the QFT formulas, it is
useful to plot the oscillating charge in time for sample values of the
masses and for different values of the momentum (we use same units for
masses and momentum). It is evident how the effect of the extra
oscillating term is maximal at lower momenta (see Figs.2 and 3) and
disappears for large $k$ (see Fig.4) where the standard oscillation
pattern is recovered. In the following plots we use $T_k=\frac{4
\pi}{\omega_{k,2} - \omega_{k,1}}$ and assume maximal mixing.

\vspace{.5cm} \centerline{\epsfysize=3.0truein\epsfbox{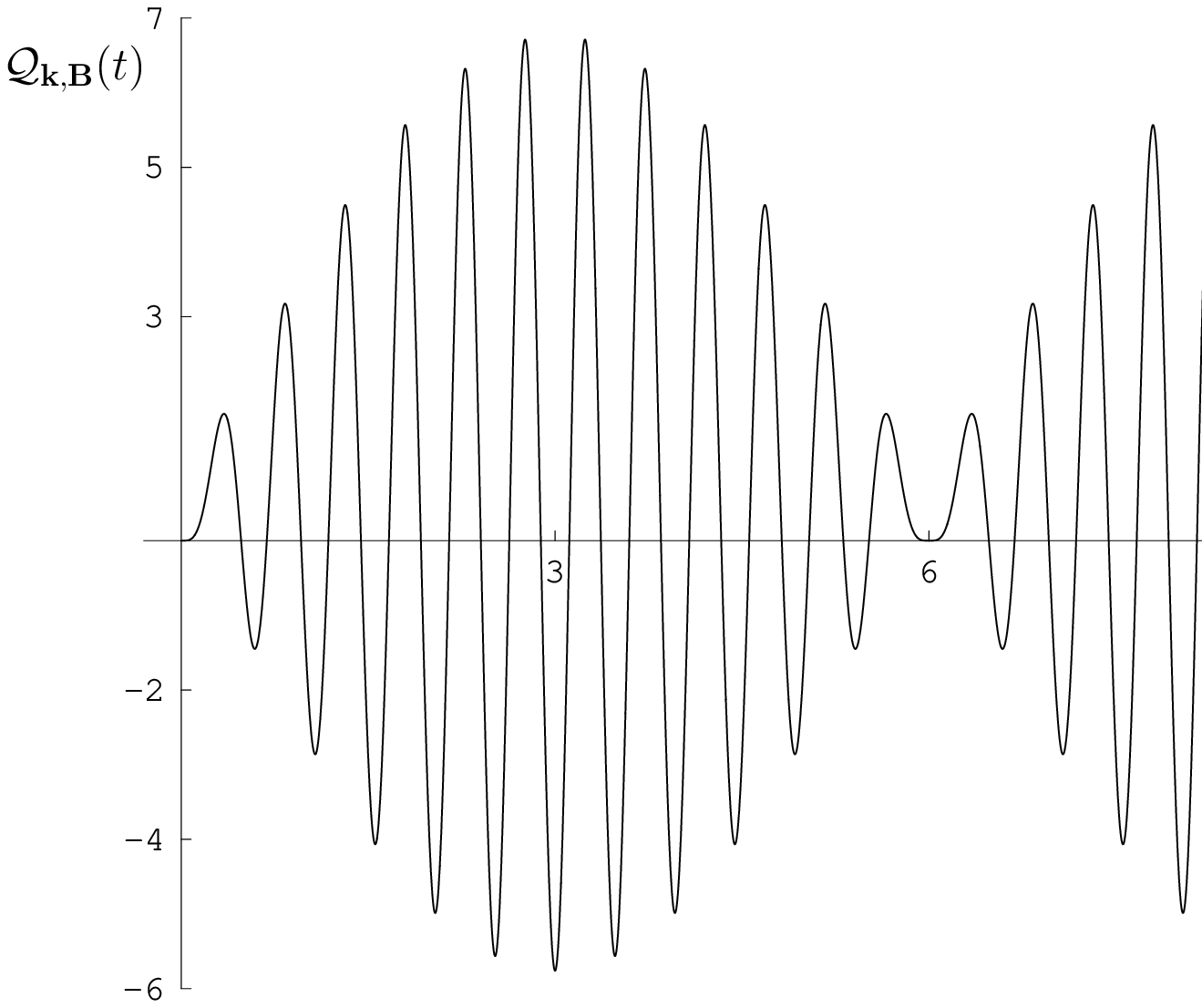}}
\vspace{.2cm} \centerline{\small Figure 2: Plot of ${\cal Q}_{{\bf
k},B}(t)$ in function of time for $k =0$, $m_1=2$, $m_2=50$ and
$\te=\pi/4$.}

\vspace{0.5cm} \centerline{\epsfysize=3.0truein\epsfbox{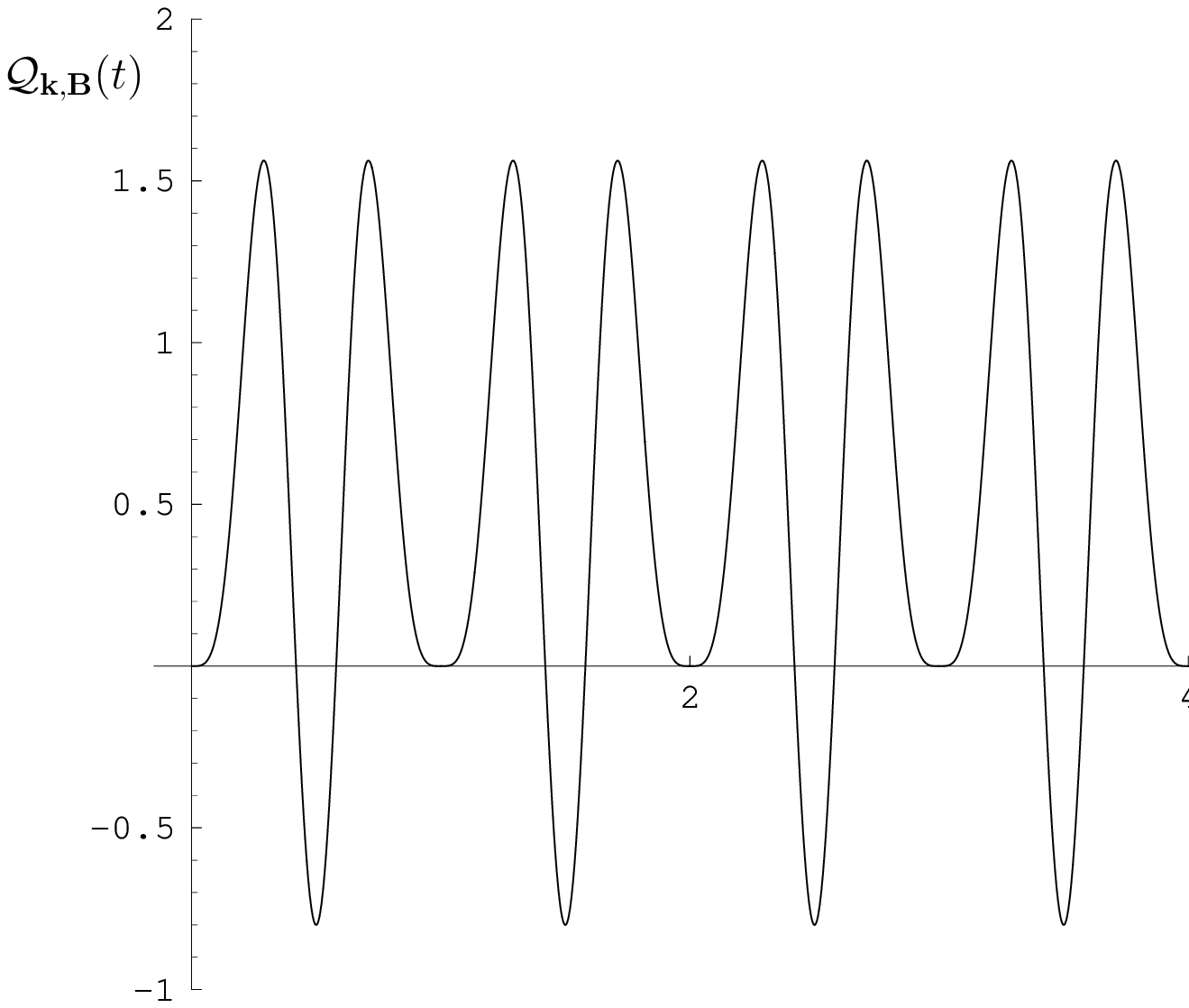}}
\vspace{.2cm} \centerline{\small Figure 3:  Plot of ${\cal Q}_{{\bf
k},B}(t)$ in function of time for $k =10$, $m_1=2$, $m_2=50$ and
$\te=\pi/4$. }

\vspace{1cm} \centerline{\epsfysize=3.0truein\epsfbox{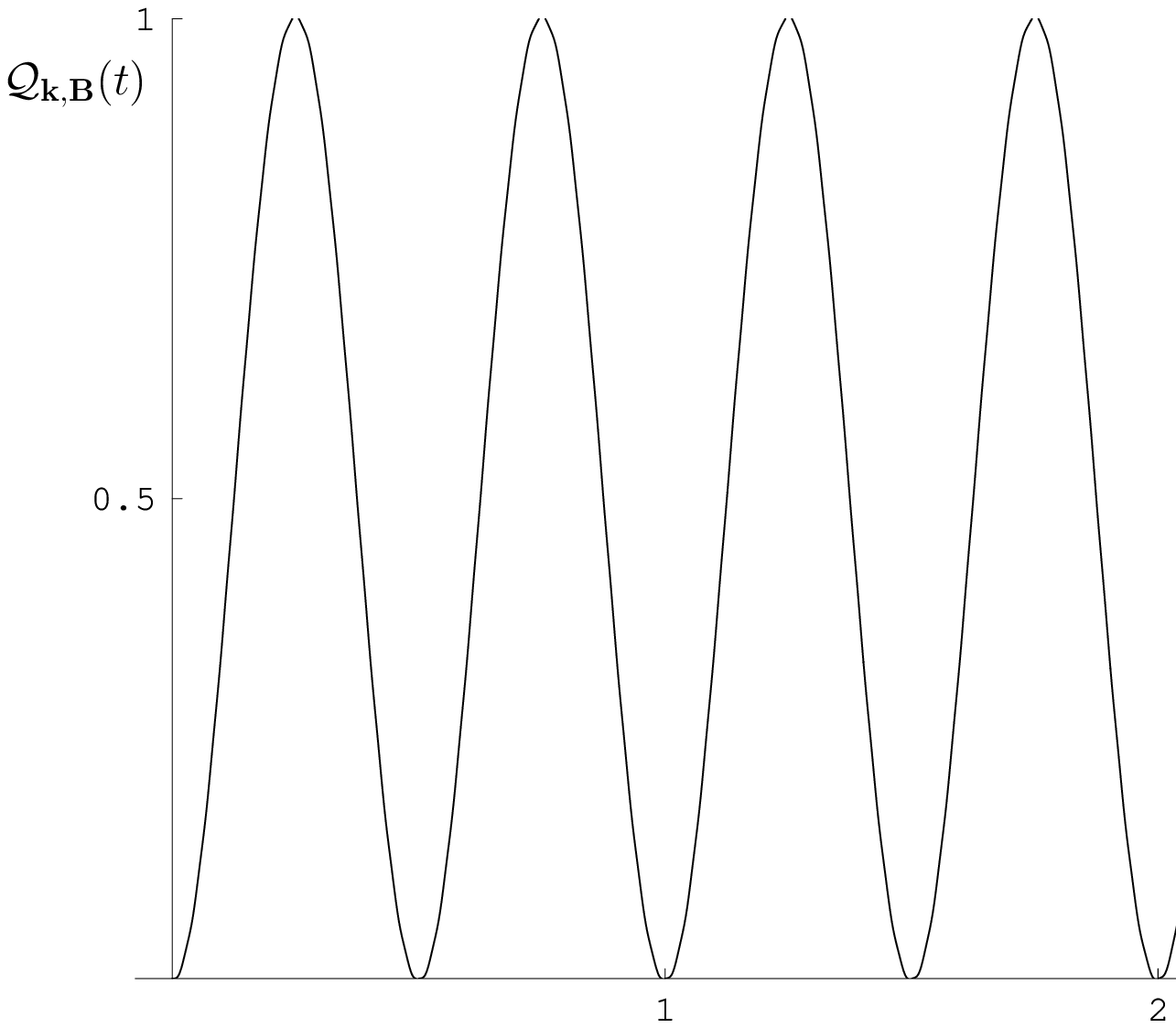}}
\vspace{.2cm} \centerline{\small Figure 4:  Plot of ${\cal Q}_{{\bf
k},B}(t)$ in function of time for $k =100$, $m_1=2$, $m_2=50$ and
$\te=\pi/4$. }

\vspace{0.5cm}

It is also interesting  to plot  the time average
of the oscillating charge,  ${\bar {\cal Q}}_{{\bf k},B} =
\frac{1}{n T_k}\int \limits_0^{n T_k} dt \,{\cal Q}_{{\bf
k},B}(t) $,
as a function of the momentum. In Figure 5 we
plot  ${\cal Q}_{{\bf k},B}(t)$ averaged over two different time
intervals, i.e. for $n=10$ and $n=100$:
it is interesting to observe how the larger is the time
interval, the more the curve converges to the average of the standard
formula, which has the value $\frac{1}{2}$.
The behavior for large k is due to the fact that, as already observed,
the exact oscillation formula reduces to the quantum
mechanical oscillation one in the large momentum limit (i.e. for
$|k|^2\gg \frac{m_1^2 +m_2^2}{2}$).

\vspace{.5cm}
\centerline{\epsfysize=3.0truein\epsfbox{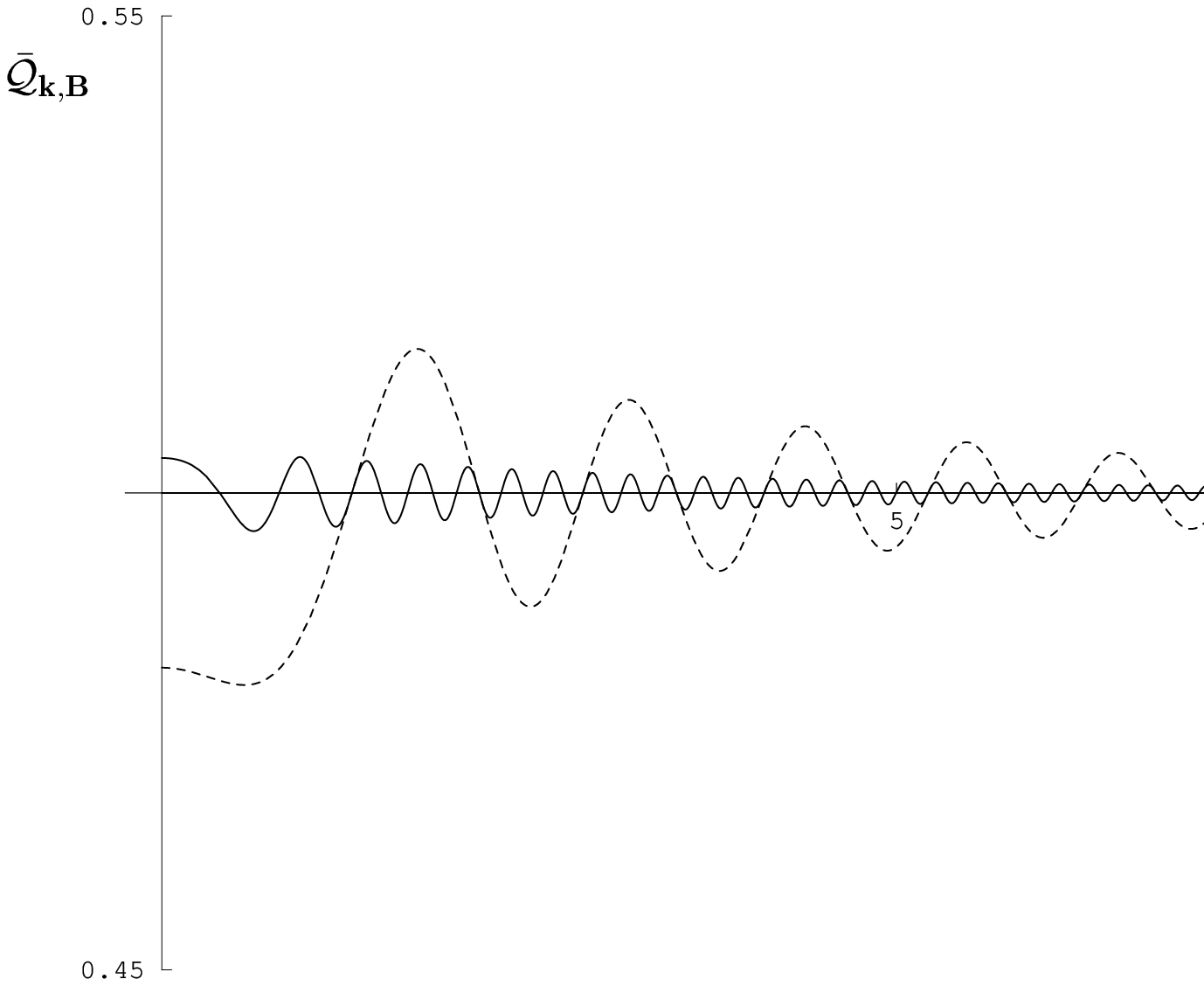}}
\vspace{.2cm}
\centerline{\small Figure 5: The time average of ${\cal
Q}_{{\bf k},B}(t)$ over 10 $T_k$  (dashed line) and over 100 $T_k$
(solid line) with respect to the average}
\centerline{\small  of the standard
oscillation formula (horizontal axis)
as a  function of $k$ for the values $m_1=2$, $m_2=50$
and  $\te=\pi/4$.}

\vspace{0.5cm}

%%%%%%%%%%%%%%%%%%%%%%%%%%%%%%%%%%%%%%%%%%%%%%%%%%%%%%%%%%%%%%%%%%
\section{Conclusions}

In this paper we have considered the quantum field theoretical
formulation of spin-zero boson field  mixing and obtained the exact
oscillation formula which does not depend on arbitrary mass parameters
which can be introduced in full generality in the theory.
We have also studied the structure of the currents and charges for the
mixed fields. In order to make our discussion more transparent, we
neglected the instability of the oscillating particles. This does
not affect the general validity of our result which rests on the
intrinsic features of QFT.

A crucial point in our analysis is the disclosure of the
fact that the space for the mixed field states is unitarily
inequivalent to the state space where the unmixed field operators are
defined. This is a common feature with the QFT structure of the fermion
mixing, which has recently been established
\cite{BV95,hannabus,Fujii:1999xa,BHV98,Blasone:1999jb}). The vacuum for the
mixed fields turns out to be a generalized $SU(2)$ coherent state. Of
course, in the boson case the condensate structure for the ``flavor''
vacuum is found to be very much different from the one in the fermion
case. Besides the intrinsic mathematical interest, our analysis
provides interesting phenomenological insights. It leads to the exact
oscillation formula for bosons which predicts oscillation behaviors
susceptible of being experimentally tested.

In fact, in the framework of the QFT analysis of Refs.
\cite{BV95,lathuile}, a study of the meson mixing and oscillations has
been already carried out in Ref.\cite{binger} and applied to the
$\eta-\eta'$ oscillation. However, the results of Ref.\cite{binger}
give observable quantities which are dependent on arbitrary mass
parameters, and this is of course physically not acceptable, as
observed in Ref. \cite{Fujii:1999xa}. In the present paper we have
pointed out the origin of such a pathology and have shown how to obtain
results which are independent from arbitrary parameters. The
oscillation formula obtained  in Ref.\cite{binger} has to be actually
replaced with the exact one here presented. In order to compare our
results with the ones of Ref. \cite{binger},  we presented in Fig.6  a
plot of charge oscillations for $\eta-\eta'$ at zero momentum, in
correspondence to what has been done in Ref.\cite{binger}.

\vspace{.5cm}
\centerline{\epsfysize=3.0truein\epsfbox{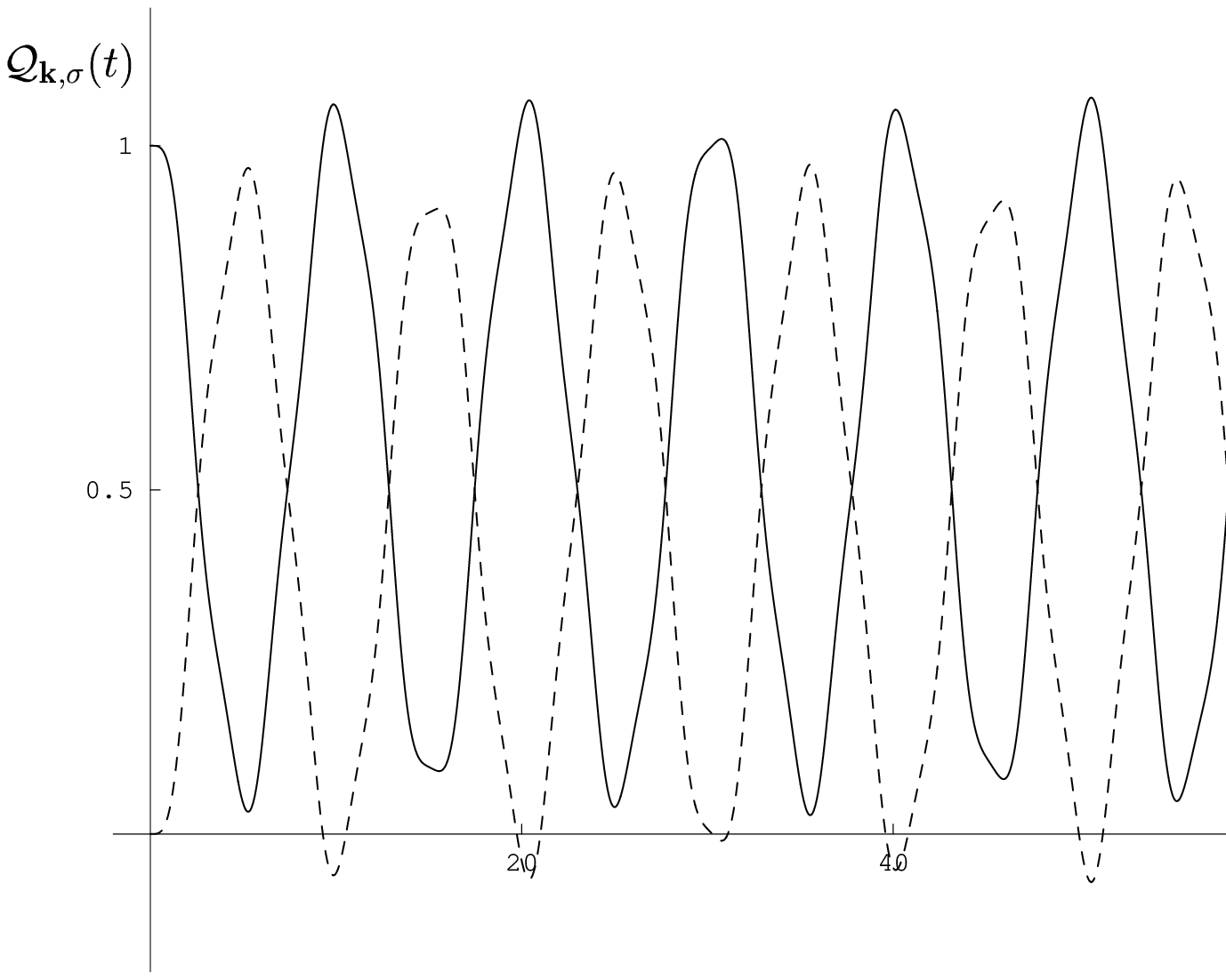}}
\vspace{.2cm}
\centerline{\small Figure 6:  Plot of ${\cal Q}_{{\bf
k},\eta}(t)$ (solid line) and ${\cal Q}_{{\bf k},\eta'}(t)$
(dashed line) in function of time }
 \centerline{\small for an initially pure $\eta$ state and
 for $k =0$, $m_1=549$ MeV, $m_2=958$ MeV and  $\te=-54^{o}$. }

\vspace{0.5cm}

Let us close by observing that although our QFT analysis discloses
features which cannot be ignored in any further study of the field
mixing and oscillations, since they are intrinsic to the structure of
the QFT formalism, nevertheless there are many aspects which are not
fully understood and many features of the physics of mixing are still
obscure \cite{zralek}, as already observed in the introduction.
The mixing of
neutrinos and their oscillations seem to be now experimentally
established and quark mixing and meson mixing are widely accepted and
verified. However, several questions
\cite{grossman,lipkin,srivastava1,lowe,kiers,burkhardt,giunti,Takeuchi:1999as}
are the object of active discussion in the framework of the quantum
mechanics formalism for neutrino oscillations.
As a matter of fact, such a state of affairs has been
a strong motivation for our searching in the structural aspects of QFT
a possible hint to the understanding of particle mixing and
oscillations.

\section{Acknowledgments}
 We thank MURST, INFN and ESF for partial support.

\vspace{.5cm}
%%%%%%%%%%%%%%%%%%%%%%%%%%%%%%%%%%%%%%%%%%%%%%%%%%%%%%%%%%%%%%%%%%%
\subsection*{Appendix: Orthogonality between mass and flavor vacua}

We calculate here  $_{1,2}\langle 0| 0(\te , t)\rangle _\AB\ $.
In the following we work at finite volume (discrete $k$) and
suppress the time dependence of the operators
when $t=0$. Let us
first observe that $| 0(\te , t)\ran_\AB= e^{i H t}|0(\te ,0)\ran_\AB$,
with $H=\sum\limits_{i=1}^2 \sum\limits_{\bf k}\, \om_{ k,i}\left(
a_{{\bf k},i}^{\dagger }a_{{\bf k},i}+b_{-{\bf k},i}^{\dagger }b_{-
{\bf k},i}\right)$. Thus we have $_{1,2}\langle 0| 0(\te , t)\rangle
_\AB\,= \, _{1,2}\langle 0| 0(\te , 0)\rangle _\AB$.

We then define $\prod\limits_{\bf k}f_0^{\bf k}(\te) \equiv \prod
\limits_{\bf k} \, _{1,2}\lan 0|  G_{{\bf k},\te}^{-1}(0)|0\ran_{1,2}$
and observe that
\bea
\frac{d}{d\te} f_0^{\bf k}(\te) &=& |V_{\bf k}| \,
_{1,2}\lan 0| (b_{{-\bf k},1} a_{{\bf k},2} +b_{{-\bf k},2}
a_{{\bf k},1} )G_{{\bf k},\te}^{-1} |0\ran_{1,2}
\\
&=& - |V_{\bf k}| \, _{1,2}\lan 0|G_{{\bf k},\te}^{-1}
( a_{{\bf k},2}^\dag
b_{{-\bf  k},1}^\dag  + a_{{\bf k},1}^\dag b_{{-\bf k},2}^\dag)
|0\ran_{1,2} ~,
\eea
where, we recall,  $|V_{\bf k}| \equiv  V_{\bf k}(0)$  in our
notation of Section II. We now consider the identity
\bea \non
(b_{{-\bf k},1} a_{{\bf k},2} +b_{{-\bf k},2}
a_{{\bf k},1})G_{{\bf k},\te}^{-1} &=& G_{{\bf k},\te}^{-1} G_{{\bf
k},-\te}^{-1} (b_{{-\bf k},1}
a_{{\bf k},2} +b_{{-\bf k},2} a_{{\bf k},1}) G_{{\bf k},-\te}
\\ \non
&=&
G_{{\bf k},\te}^{-1}[ b_{{-\bf k},A}(-\te) a_{{\bf k},B}(-\te) +
b_{{-\bf k},B}(-\te) a_{{\bf k},A}(-\te)] ~.
\eea
Then the equation follows
\bea
\frac{d}{d\te} f_0^{\bf k}(\te) &=& -2\, |V_{\bf k}|^2
\cos\te \sin\te  f_0^{\bf k}(\te) +\sin^{2}\te \,|V_{\bf k}|^3 \,
_{1,2}\lan 0|G_{{\bf k},\te}^{-1} ( a_{{\bf k},2}^\dag b_{{-\bf
k},1}^\dag + a_{{\bf k},1}^\dag b_{{-\bf k},2}^\dag) |0\ran_{1,2}
\\
&=&  -2 \,|V_{\bf k}|^{2} \cos\te \sin\te f_0^{\bf k}(\te ) -
\sin^{2}\te |V_{\bf k}|^{2} \frac{d}{d\te} f_0^{\bf k}(\te)
\eea
and
\bea
\frac{d}{d\te} f_0^{\bf k}(\te) &=& -\frac{2 |V_{\bf k}|^{2} \cos\te
\sin\te }{1+\sin^{2}\te |V_{\bf k}|^{2}}  f_0^{\bf k}(\te) ~,
\eea
which is solved by
\bea\label{A74}
f_0^{\bf k}(\te) &=& \frac{1 }{1+\sin^{2}\te |V_{\bf k}|^{2}}
~,
\eea
with the initial condition $f_0^{\bf k}(0)=1$.

We observe that we can  operate in a similar fashion  directly
with $f_0^{\bf k}(\te , t) \equiv \, _{1,2}\lan 0|  G_{{\bf
k},\te}^{-1}(t)|0\ran_{1,2}$. We then find that $f_0^{\bf k}(\te ,
t)$ is again given by Eq.(\ref{A74}) and thus it is actually
time-independent.
We also note that by a similar procedure it can be proved that
 $\lim\limits_{V\rar\infty}\, {}_\AB\langle
0(\te,t)|0(\te',t)\ran_\AB \rar 0$ for $\te'\neq\te$.

%%%%%%%%%%%%%%%%%%%%%%%%%%%%%%%%%%%%%%%%%%%%

%%%%%%%%%%%%%%%%%%%%%%%%%%%%%%%%%%%%%%%%%%%%%%%%%%%%%%%%%%
%

\end{document}